\documentclass[12pt]{elsart}
\usepackage{graphics,epsfig}
\begin{document}
\begin{frontmatter}
\title{A phenomenological analysis of 
antiproton interactions at low energies} 
\author[Brescia]{A.~Bianconi},
\author[Brescia]{G.~Bonomi},
\author[Brescia]{M.P.~Bussa},
\author[Brescia]{E.~Lodi Rizzini},
\author[Brescia]{L.~Venturelli},
\author[Brescia]{A.~Zenoni},
\address[Brescia]{Dip. di Chimica e Fisica per 
l'Ingegneria e per i Materiali, 
Universit\`{a} di Brescia and
INFN, Sez. di Pavia, Italy} 

\begin{abstract}
We present an optical potential analysis of the $\bar{p}p$ 
interactions at low energies. Our optical potential is purely 
phenomenological, and has been parametrized on data recently 
obtained by the Obelix Collaboration
at momenta below 180 MeV/c. It reasonably 
fits annihilation and elastic data below 600 MeV/c, and allows 
us for an evaluation of the elastic cross section  
and $\rho-$parameter down to 
zero kinetic energy. Moreover we show that 
the mechanism that depresses 
$\bar{p}-$nucleus annihilation cross sections at low 
energies is present in $\bar{p}p$ interactions too. 
\end{abstract} 
\end{frontmatter} 

\section{Introduction} 

Recently data on $\bar{p}$ annihilations on 
light nuclei (H, D and $^4$He) 
have become available at very small $\bar{p}$ momenta (down to 
45 MeV/c)\cite{obe1,obe2,obe3,obe4}. 
A new data on 
$^{20}$Ne at 57 MeV/c is also available now\cite{ne991}. 
Together with 
previously available data (for a review see 
e.g. ref.\cite{bende}), and with data on antiprotonic 
atoms\cite{bat90,wid}, the full set presents some 
interesting features, that we will try and correlate in 
this work. As far as a qualitative physical understanding 
is concerned, the unifying feature is a mechanism that we  
call ``inversion'', i.e. a repulsion-dominated 
low energy $\bar{p}p$ interaction. 
From a practical point of view, we will 
widely rely on 
the possibility of reproducing the available elastic 
and annihilation $\bar{p}p$ data below 600 MeV/c 
via an energy independent optical potential. 

Let us initially discuss some 
relevant points of the phenomenology: 

1) Annihilation $\bar{p}p$ 
data show, in a log-log plot, a series of 
roughly rectilinear behaviors (see fig.1). 
These can be approximately identified 
with regions where different angular momentum 
components are dominant, with the S-P transition 
at about 100 MeV/c. At 50 MeV/c it is possible to 
assume S-wave dominance and estimate the imaginary part 
of the scattering length $\alpha$\cite{pro2}. The real part, 
is extracted from the widths and shifts of 
the levels of antiprotonic Hydrogen atoms\cite{bat90},
together with an independent measure of $Im(\alpha)$.  
Elastic $\bar{p}p$ data at values of the laboratory $\bar{p}$ 
momentum $k$ in the 200-500 MeV/c range were 
reproduced by Br\"uckner $et$ $al$\cite{bru86} 
with a phenomenological optical potential. These authors 
left a wide range of uncertainty for the suggested 
potential parameters. We have noticed 
that the same potential, with a finer tuning of 
the parameters, can fit all the annihilation 
data which have been later 
measured at smaller $k$, down to 30 MeV/c, 
by the Obelix Collaboration\cite{obe1,obe3,obe4} 
(see fig.2, and section 3 for details). It can also 
calculate the real and imaginary parts of the 
scattering length. Optical potential analysis, partial 
wave analysis and atomic data 
agree on $Re(\alpha)$ $\approx$ 
$-Im(\alpha)$ $\approx$ 0.7$\div$0.8 fm, with positive 
sign. 

\begin{figure}[htp]
\begin{center}
\mbox{
\epsfig{file=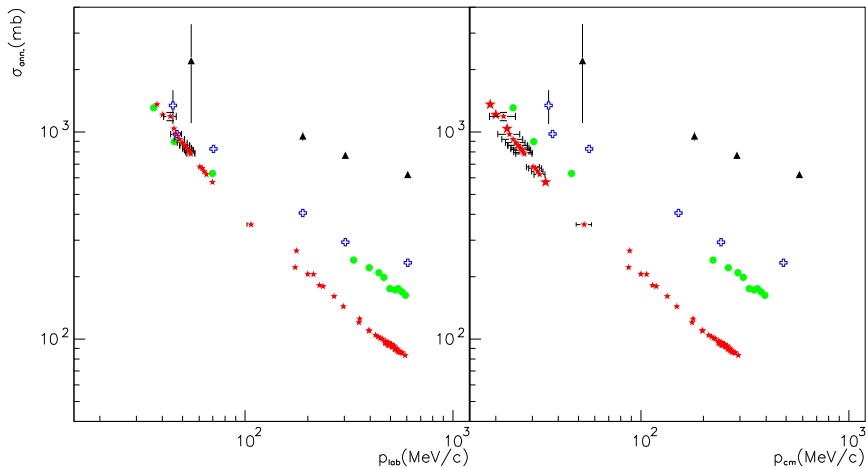,width=0.9\linewidth}}
\end{center}
\caption[]
{\small\it \label{fig1}
Annihilation cross sections (mb) vs laboratory 
(left) and center of mass (right) momentum. 
Stars correspond to $\bar{p}p$\cite{obe1,obe3,obe4,bru90} 
circles to $\bar{p}D$\cite{obe2,deu}, crosses to 
$\bar{p}^4$He\cite{obe2,elio1,elio2} and triangles to 
$\bar{p}^{20}$Ne\cite{neon,ne991}. }
\end{figure}

2) The $\rho$ parameter, i.e. the  ratio between the 
real and imaginary part of the forward scattering amplitude, 
can be measured at zero or near zero energy 
exploiting $\rho(0)$ $=$ $Re(\alpha)/Im(\alpha)$, which 
means $\rho$ $\approx$ $-1$. At larger energy, it 
must be extracted by a $very$ delicate (and partly model 
dependent) analysis of the elastic $\bar{p}p$ angular 
distributions. 
Despite the behavior of $\rho$ is still unclear in 
the region 100-200 MeV/c\cite{dst94}, an overview of 
the experimental 
data\cite{bru85,cre,iwa,ash,lin,bat90} 
suggest that $\rho$ is small but positive (0.1$\div$0.3) 
at projectile momenta over some value which lies somewhere 
around 500 MeV/c, smaller 
(with uncertain sign) in the region 180-500 MeV/c,  
and tends to some negative value $\sim$ $-1$ at zero energy. 
As better described in the following, we have applied 
the optical potential (whose parameters have been fine-tuned 
on the $\bar{p}p$ annihilation data at 30-100 MeV/c), 
to predict the $\rho$ behavior. The results 
agree with the large and the zero energy data, and 
suggest that $\rho$ varies monotonously in the less known 
intermediate momentum region (see fig.3). 

3) In the laboratory frame 
$\bar{p}$ total annihilation cross sections 
(TPA from now on) on Deuteron and $^4$He 
are almost equal, and both are smaller than TPA on Hydrogen.  
The $^{20}$Ne datum is larger, but not 
so large as one could expect\cite{ne991}. See fig.1 for 
a general 
view of the data. Taking into account that a large enhance 
of reaction cross sections is predicted 
at low energies because of  
charge effects\cite{ll1}, this phenomenon is surprising. 
According with the notations 
used in previous works\cite{noi1,noi2} we will 
call this behavior ``inversion''. 
Actually, if the data are represented in the center of mass 
frame, TPA on D and $^4$He are slightly larger than 
TPA on Hydrogen, however the dependence of the 
TPA on the mass number is still much smaller than any 
geometrical expectance (see later for a discussion 
of the ``geometrical expectance'' and of the role of 
the center of mass). For 
$k$ $>>$ 100 MeV/c this phenomenon is 
not observed and the ratio between $\bar{p}p$ and 
$\bar{p}^4$He annihilation 
rates is qualitatively what one would expect. The 
inversion behavior  
is confirmed by an analysis of antiprotonic atoms\cite{wid}, 
where it is found that $\vert Im(\alpha) \vert$ 
is smaller in antiprotonic Deuterium than in antiprotonic 
Hydrogen. 

4) From an overview of the available 
$\bar{p}-nucleus$\cite{bende} and 
$\bar{n}-nucleus$\cite{ableev} annihilation data below 600 MeV/c 
it appears that: (i) Where many partial waves dominate the 
$\bar{p}-$nucleus interaction the cross sections 
relative to different 
nuclear species are parallel, and agree with a law 
$\sigma$ $\propto$ $\sigma_oA^{2/3}$. (ii) Where only a few 
partial waves are supposed to dominate, a convergency 
(for decreasing energies) between the  different TPA 
is clearly visible. In a log-log plot, the extrapolations 
of the different TPA seem to aim at some common intersection 
point somewhere at $k_{cm}$ $\sim$ 1 MeV/c  
(see fig.1). 

\section{General theoretical background.} 

To better understand the significance of the previous 
nuclear data some considerations are useful. 
Both the ``inversion'' and the convergency behavior 
contraddict the geometrical predictions.  
Assuming that the imaginary 
part of the scattering length is roughly equal 
to the nuclear size $R$ $\approx$ 
1.3$A^{1/3}$ fm, and exploiting the 
traditional estimation of the Coulomb 
focusing effect\cite{ll1,cph}, one 
has TPA $\sim$ $ZA^{1/3}/k^2$ at very small 
momenta. At larger momenta the semiclassical  
expectance is TPA $\propto$ $A^{2/3}$ (well verified 
for $\bar{p}$\cite{bende} and $\bar{n}$\cite{ableev} 
annihilations at any 
$k_{lab}$ $>$ 180 MeV/c). Since for most nuclei 
$ZA^{1/3}$ $\approx$ $0.5 A^{4/3}$, one should naively 
expect that TPA on different nuclei increase their separation 
when momenta decrease below 100 MeV/c, while exactly the 
opposite in seen. In addition, at 
any precise lab or c.m. momentum below $k_{lab}$ $=$ 
100 MeV/c, the $A-$dependence of the known TPA is below 
both the $A^{2/3}$ and the $ZA^{1/3}$ prediction. 

Regarding the question whether the TPA on different nuclei 
must be compared at the same laboratory or center of mass 
momenta, the answer is model-dependent. 
In Impulse Approximation inspired models, the annihilation 
process only involves one of the nucleons in the target 
nucleus, which has average momentum equal to zero in 
the laboratory. It is then reasonable 
to compare data at the same laboratory momentum. 
In compound-nucleus inspired models, the collision 
process directly transfers momentum from the projectile to 
the full target. In this case, data taken on different 
targets should be compared at the same c.m. momentum. 

\begin{figure}[htp]
\begin{center}
\mbox{
\epsfig{file=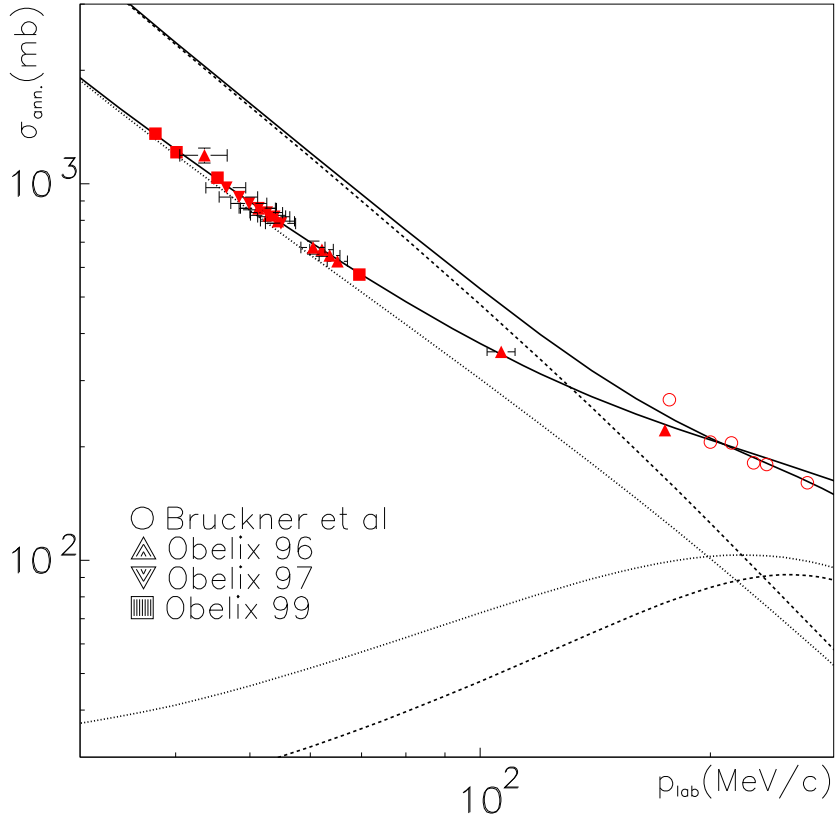,width=0.9\linewidth}}
\end{center}
\caption[]
{\small\it \label{fig2}
Optical potential fits to $\bar{p}p$ annihilation data. 
The continuous line fitting 
the low energy points corresponds to the potential 
described in the text. The upper continuous line 
corresponds to the same potential modified by 
decreasing the imaginary strength from 8000 to 1000 MeV. 
Dotted lines show the S- and P-wave contributions for 
the former potential, dashed lines show the same for the 
latter. Data are taken from Br\"uckner $et$ $al$\cite{bru90},
and from the Obelix collaboration\cite{obe1,obe3,obe4}.}  
\end{figure}

The key point is the generalization of the 
concept of low energy ``inversion''. 
On the ground of general quantum principles 
it is possible to demonstrate\cite{pro1,noi1} that, 
in presence of a $very$ effective 
esothermic hadronic reaction mechanism 
and in conditions of S-wave dominance: (i) 
the reaction cross section 
must stay much below geometrical expectations, 
and is largely independent on the target nucleus 
size; (ii) most attempts to increase those model 
parameters which supposedly should enhance the 
annihilation rate (e.g. strength or radius of a potential) 
lead to the opposite or to no result; (iii) 
a strong non-diffractive elastic scattering 
accompanies the reaction at low energies, and this 
scattering has $repulsive$ character (i.e. 
$Re(\alpha)$ $>$ 0). So, with ``inversion'' 
we will refer to the presence of these three features. 
We have previously demonstrated\cite{noi1} that 
strong inversion must be expected whenever 
disappearance of the projectile S-wave 
wavefunction $\Psi_S$ (at the nuclear surface) is 
produced within a range much smaller than 
$\Delta r$ $\approx$ 1 fm. Then, 
regularity conditions on $\Psi_S$ at the nuclear surface 
necessarily produce a large flux reflection and a 
$\Psi_S$ which is similar to the one produced by a 
repulsive potential with little absorbtion. For this reason 
it is not proper to consider the scattered flux 
as ``diffractive'', although it is a by-product of 
absorbtion. It is a refractive process, as in elastic 
potential scattering. Now we can 
better specify the above required condition 
of ``very effective 
reaction mechanism'' (since at low energies it is not 
so effective): It means that (i) the reaction 
is esothermic, (ii) it produces large reaction rates 
at large energies, (iii) at any energy its free mean 
path in nuclear matter can be estimated to be shorter 
than 1 fm. We remark that the described behavior 
is experimentally confirmed by the fact that for the 
$\bar{p}p$ scattering length $\alpha$ we have 
$Re(\alpha)$ $\approx$ $-Im(\alpha)$ $>$ 0, 
or equivalently $\rho(0)$ $\approx$ $-1$. And by the fact 
that $\bar{p}$ annihilation rates on nuclei are 
not that large. 

Also the traditional view of the Coulomb focusing 
effect must be reconsidered. 
In a previous paper\cite{noi2} we have already 
calculated and compared ``charged'' and ``uncharged'' 
annihilation rates on nuclei with finite size, and 
demonstrated that the traditional $Z/\beta$ 
Coulomb enhancement factor\cite{ll1} 
is exagerated. This factor is estimated with the 
two assumptions: (i) pointlike target (ii) completely  
independent action of Coulomb and strong forces. 
On the contrary, on one side the interplay between 
Coulomb and strong forces is not negligible, 
and on the other side finite size effects largely 
neutralize 
the Coulomb enhancement factor for intermediate and 
heavy nuclear targets. E.g., speaking 
in terms of target effective charge $Z_e$, 
we have $Z_e(^4He)/Z_e(H)$ 
$\approx$ 1 (instead of 2), 
$Z_e(^{20}Ne)/Z_e(^4He)$ $\approx$ 2 
(instead of 5; comparisons are 
performed at the same laboratory momentum, but center of mass 
effects were included in the calculation\cite{noi2}). 

\section{Optical potential fits on $\bar{p}p$ data}. 

As previously anticipated, 
all the data on $\bar{p}p$ elastic and annihilation 
cross section below 600 MeV/c 
can be reasonably well fitted by  
the same potential, with Woods-Saxon shape, 
used by Br\"uckner $et$ $al$\cite{bru86} to 
fit elastic $\bar{p}p$ data at 181, 287 and 505 MeV/c, 
after a finer tuning of the parameters. We have set 
the real and imaginary strength to -46 and -8000 MeV, 
the real and imaginary radius to 1.89 and 0.41 fm, 
and the diffuseness to the common value 
0.2 fm. The fit on the annihilations is very good below 300 
MeV/c and good within 10 \% at 600 MeV/c 
(the exact precision over 300 MeV/c depends 
on which set of data is chosen\cite{bende}), and the elastic 
distributions are still well reproduced. The total 
potential includes the Coulomb potential of a 
spherical charge distribution 
with radius 1.25 fm. In all the calculations 
center of mass corrections have been included. 
Together with the outcome of the above potential, in 
fig.2 we also show a curve corresponding 
to imaginary strength 1000 MeV. For both cases (strength 
8000 and 1000 MeV) we also show the S- and P-wave 
contributions. Evidently the used potential 
does produce ``inversion'', i.e. a larger annihilation 
potential produces a smaller annihilation rate. 
From the same figure it is obvious that 
this behavior is associated with the S-wave dominance, 
and is present only below an ``inversion point'' 
$k_{inv}$. In this case $k_{inv}$ $\approx$ 200 MeV/c. 

\begin{figure}[htp]
\begin{center}
\mbox{
\epsfig{file=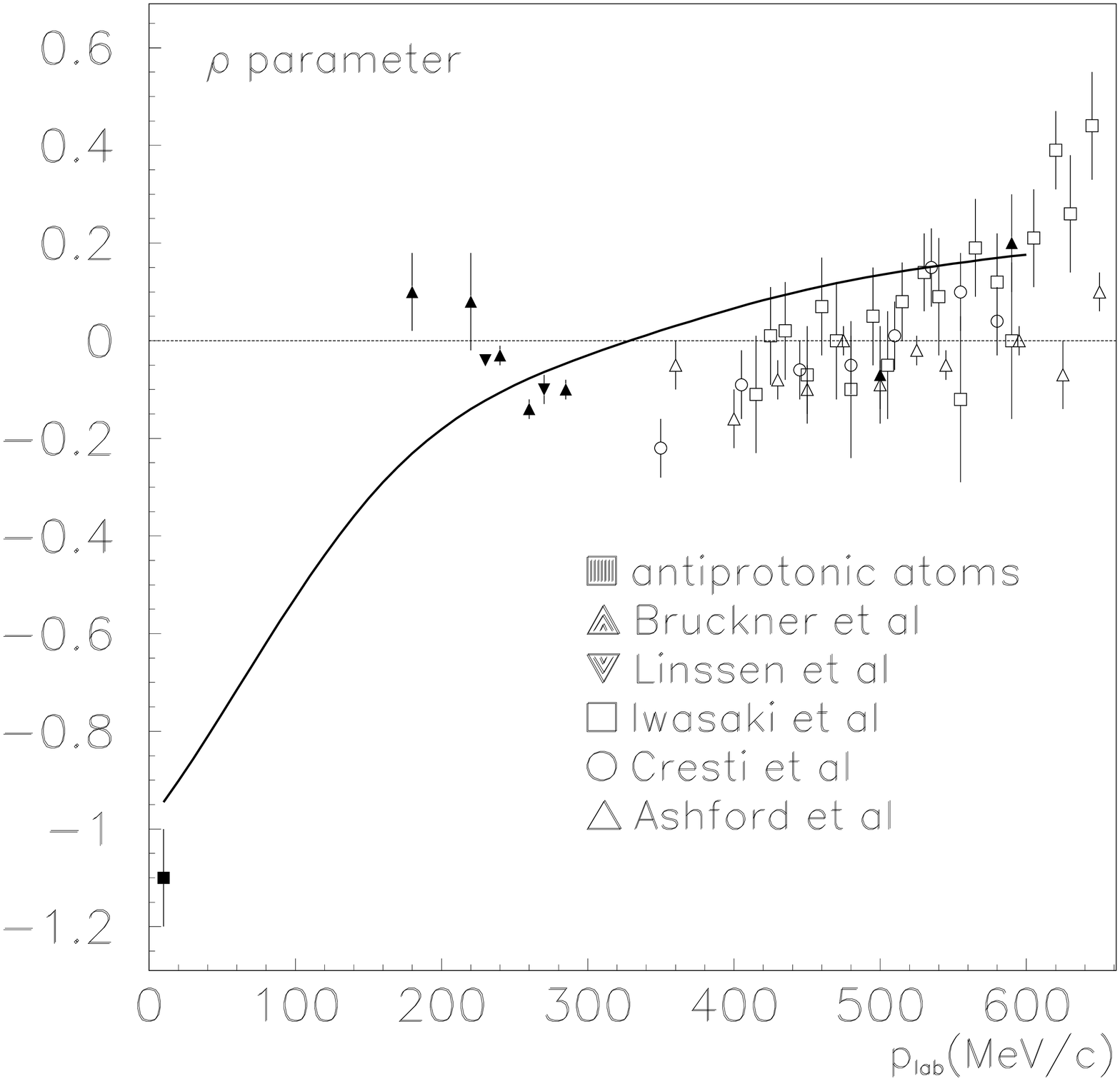,width=0.9\linewidth}}
\end{center}
\caption[]
{\small\it \label{fig3}
Continuous line: The optical potential prediction for 
$\rho$. Scattering data up to 650 MeV/c come from 
M.Cresti et al (1983)\cite{cre},
H.Iwasaki et al (1985)\cite{iwa},
V.Ashford et al (1985)\cite{ash}
W.Br\"uckner et al (1985)\cite{bru85}
L.Linssen et al (1987)\cite{lin}. For the atomic data see 
e.g. C.J.Batty et al\cite{bat90} and references therein, 
and also references contained in \cite{wid}.  
}
\end{figure}

In fig.3 we show the value of 
$\rho$ in the momentum range 0-600 MeV/c calculated with 
this potential. The change of sign of $\rho$ 
can be related with the transition from the dominance 
of the reaction-associated repulsion to the dominance of the 
direct potential attraction, at least in forward scattering. 
Indeed, at increasing momenta the Born approximation 
becomes progressively more reliable, and it permits to 
estimate $\rho$ $\sim$ $(V_R R_R^3)/(V_IR_I^3)$ $\sim$ 
$+$0.2, using as an effective radius the sum of the potential 
radius and diffuseness.  
The positive $\rho$ value at large momenta 
is thus directly due to the presence of a real attracting 
part in the potential. 
We notice that the ``source'' of the ``direct'' attraction 
will be the region where absolute value of 
the elastic potential is roughly 
equal to the kinetic energy, while the 
``source'' of the reaction-induced repulsion will be 
the region where most annihilations take place, i.e. 
0.5$\div$1 fm out of the edge of the annihilation core  
This distance has been estimated in past years in 
analysis of both 
$\bar{p}p$\cite{bru86} and $\bar{p}-$nucleus\cite{aar} 
interactions. 

\begin{figure}[htp]
\begin{center}
\mbox{
\epsfig{file=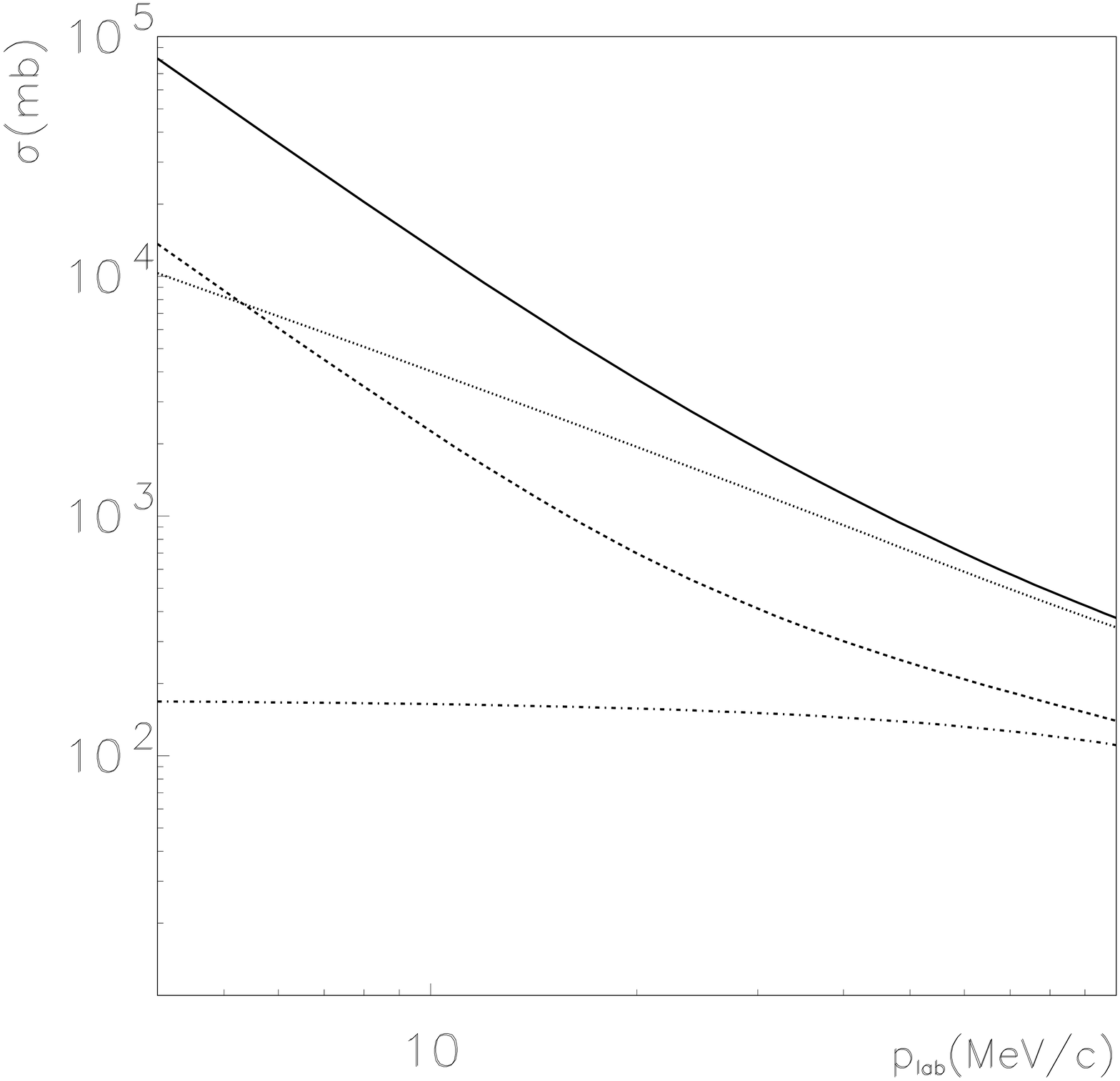,width=0.9\linewidth}}
\end{center}
\caption[]
{\small\it \label{fig4}
Continuous line: Total $\bar{p}p$ 
annihilation cross section calculated as in fig.2. Dashed 
line: Total strong elastic 
cross section (the Coulomb forward singularity is 
excluded). Dotted and dot/dashed line: Total 
annihilation and elastic cross sections calculated again 
after assuming zero charge for the projectile. At larger momenta 
the charge does not create large differences. At lower 
momenta 
$\sigma_{ch}^{ann}$ $\sim$ $1/\beta^2$, 
$\sigma_{ch}^{el}$ $\sim$ $1/\beta^2$, 
$\sigma_{neu}^{ann}$ $\sim$ $1/\beta$, 
$\sigma_{neu}^{el}$ $\sim$ constant. }
\end{figure}

In fig.4 the total annihilation and elastic cross sections 
are reported, compared with the corresponding cross sections 
calculated after turning off the electric charge.  
In the former case the contribution 
of the pure Coulomb forward peak 
and of the Coulomb-strong interference is excluded.  
Nevertheless, 
the elastic strong cross section is 
largely affected by Coulomb focusing effects. In 
particular, the figure shows that 
the ratio between the strong elastic and 
the annihilation total cross sections is completely 
dominated by the Coulomb effects. Without them, 
$\sigma_{el}/\sigma_a$ $\rightarrow$ 0 for $k$ 
$\rightarrow$ 0. With inclusion of the charge effect, 
approximately $\sigma_{el}/\sigma_a$ $\rightarrow$ 1/6.  
We have also 
calculated angular distributions at momenta between 25 
and 100 MeV/c, but they are practically flat up to 
50 MeV/c, and at 100 MeV/c present a 20\% change between 
forward and backward scattering, so they are not very 
interesting. At 100 MeV/c the P-wave contributions 
are 1 \% in the total strong elastic cross section, and  
10\% in the annihilation. We remark 
that at such small momenta the Rutherford ``forward'' 
peak, which spreads at angles $\theta$ $\propto$ $1/k$, 
becomes the most important source of 
elastic scattering at large angles too. 

\section{Annihilations on nuclei.} 

Up to now we did not succeed in fitting 
light nuclei data perfectly by energy-independent optical 
potentials (which take nuclear density 
distributions into account). In fact, at momenta below 
100 MeV/c a certain energy dependence is introduced by the 
nontrivial energy dependence of the 
$\bar{p}n$ annihilation rate\cite{feli,arm}). 
The study of the nuclear optical potential 
requires taking into account nuclear structure details 
and $\bar{p}n$ interactions, so a more specific and 
longer work will be devoted to it in the next future. 
Qualitatively, it is 
evident that the energy dependence of the cross sections 
in the range 30-200 MeV/c 
is much slower in $\bar{p}-$nucleus than in $\bar{p}p$.  
This can be related to the change of sign of $\rho$ in 
$\bar{p}p$ interactions observing that if the  
$\bar{p}-$nucleon interaction is repulsive below a 
certain momentum of scale $\sim$ 100 MeV/c, in a cluster 
of nucleons each single nucleon will contribute keeping 
the projectile far from itself and from all the other 
ones. In the language of the multiple 
scattering expansion this is an interference between single 
and double scattering processes, i.e. elastic 
scattering of $\bar{p}$ on one nucleon prevents annihilation on 
another one. This interpretations would confirm the 
suggestion given by Wycech $et$ $al$ in their analysis 
of antiprotonic deuterium\cite{wyc}: they estimate single 
and double scattering amplitudes contributing to the 
$\bar{p}D$ interaction, and observe that the interference 
between them decreases the single scattering output. 
At the same time our calculations (still in progress) show that, 
in the case of light nuclei, nuclear structure details  
and $\bar{p}n$ features do affect the 
results. 

\section{Conclusions.} 

We have shown that the Obelix Collaboration data on 
$\bar{p}p$ annihilation in the range 30 to 180 MeV/c 
allow us for a finer tuning of the parameters of 
an optical potential, which was previously used by 
other authors to interpolate elastic 
differential cross sections at $k_{lab}$ 181, 287 
and 505 MeV/c. Without the need of introducing any 
energy dependence of these parameters, the 
so-obtained potential can 
reproduce all the $\bar{p}p$ 
annihilation data between 30 and 600 MeV/c, 
the zero-energy value 
of the $\rho$ parameter together with its general trend at 
increasing energies, and the measured values of the scattering 
length (real and imaginary part) with correct sign. 
We have also used this potential to predict elastic 
cross sections and $\rho$ values in those regions where 
data are not available yet. 
We have also shown that the behavior of all the considered 
observables is largely affected by a mechanism that 
we call ``inversion'': in presence of a very strong 
reaction mechanism the reaction cross sections become 
anomalously small at very low energies, while 
elastic interactions  reverse from attractive to repulsive.  
We can't make precise predictions for the 
$\bar{p}-$nucleus cross sections yet, but we stress that 
their smallness is closely related with the low-energy 
repulsive behavior of the $\bar{p}p$ interaction.

\end{document}